\RequirePackage{fixltx2e}
\documentclass[runningheads,a4paper]{llncs}

\usepackage[american]{babel}

\usepackage{graphicx}

\usepackage{paralist}

\usepackage{cite}

\usepackage{csquotes}

\usepackage[T1]{fontenc}

\usepackage{lmodern}

\usepackage{microtype}
\usepackage{color}
\usepackage{graphicx}
\usepackage{subfigure}
\usepackage{amsmath}
\usepackage{amssymb}
\usepackage{graphicx}
\usepackage{subfigure}
\usepackage{algorithm}
\usepackage{algpseudocode}
\usepackage{color}
\usepackage{todo}
\usepackage{color}
\usepackage{xcolor}
\usepackage{multirow}

\ifnum\pdfoutput>0
\usepackage[
bookmarks=false,
breaklinks=true,
colorlinks=true,
linkcolor=black,
citecolor=black,
urlcolor=black,
pdfpagelayout=SinglePage
]{hyperref}
\usepackage[all]{hypcap}
\else
\usepackage{hyperref}
\fi

\usepackage[capitalise]{cleveref}
\crefname{section}{Sect.}{Sect.}
\Crefname{section}{Section}{Sections}
\crefname{figure}{Fig.}{Fig.}
\Crefname{figure}{Figure}{Figures}

\usepackage{xspace}
\newcommand{\eg}{e.\,g.,\ }
\newcommand{\ie}{i.\,e.,\ }
\newcommand{\ra}{\rightarrow}

\newcommand{\ACRONYM}{\texttt{NETCS}\xspace}
\newcommand{\gring}{\textsf{GlobalRing}\xspace}
\newcommand{\ccover}{\textsf{CycleCover}\xspace}
\newcommand{\gline}{\textsf{GlobalLine}\xspace}
\newcommand{\gstar}{\textsf{GlobalStar}\xspace}
\newcommand{\dprime}{{\prime\prime\xspace}}

\begin{document}

\pdfgentounicode=1

\title{\ACRONYM: A New Simulator of Population Protocols and Network Constructors 
\thanks{Supported in part by the project ``Foundations of Dynamic Distributed Computing Systems'' (\textsf{FOCUS}) which is implemented under the ``ARISTEIA'' Action of the  Operational Programme ``Education and Lifelong Learning'' and is co-funded by the European Union (European Social Fund) and Greek National Resources.}
}


%
\iftrue
\author{Dimitrios Amaxilatis\inst{1} \and Marios Logaras\inst{1} \and Othon Michail\inst{1} \and Paul G. Spirakis\inst{1, 2}}

\institute{
Computer Technology Institute $\&$ Press ``Diophantus'' (CTI), Patras, Greece\\
\email{ {\{amaxilat,logaras,michailo\}@cti.gr} }\and
Department of Computer Science, University of Liverpool, UK\\
\email{P.Spirakis@liverpool.ac.uk}
}
\fi
			
\maketitle

\begin{abstract}
Network Constructors are an extension of the standard population protocol model in which
finite-state agents interact in pairs under the control of an adversary scheduler.
In this work we present \ACRONYM, a simulator designed to evaluate the performance of various network constructors and population protocols under different schedulers and network configurations.
Our simulator provides researchers with an intuitive user interface and a quick  experimentation environment to evaluate their work.
It also harnesses the power of the cloud, as experiments are executed remotely and scheduled through the web interface provided.
To prove the validity and quality of our simulator we provide an extensive evaluation of multiple protocols with more than \emph{100000} experiments for different network sizes and configurations that validate the correctness of the theoretical analysis of existing protocols and estimate the real values of the hidden asymptotic coefficients.
We also show experimentally (with more than  \emph{40000} experiments) that a probabilistic algorithm is capable of counting the actual size of the network in bounded time given a unique leader.
\end{abstract}

\keywords{distributed network construction, distributed protocol simulation, network simulation, random schedule, probabilistic counting}

\section{Introduction}\label{sec:intro}

Network Constructors~\cite{MS14} are an extension of the standard population protocol model~\cite{AADFP06} in which
finite-state agents interact in pairs under the control of an adversary scheduler.
The automata (also called nodes) reside in a well-defined area without being capable of moving or change the form of the network.
One of the main characteristics of such networks is the limited resources both in processing power and storage capacity.
However, the nodes can cooperate by interacting in pairs.
Every such interaction may result in an update of the local states of the nodes or communication links to form the desired configuration.
Their basis of their behavior is of a dynamic distributed computing system with applications to the operation of computers inside the world wide web, processes inside an operating system or mobile phones using technologies like Bluetooth or Nfc to exchange information in a distributed manner.

In this work, we present an experimentation platform that simulates the behavior of such automata for any given protocols and under different conditions or network sizes.
One of the main novelties of our work, concerns the ease of testing and validating assumptions with minimal effort and hassle to help identify errors and better understand the behavior of protocols in different scenarios and under different schedulers according to the overall model.
To highlight the advantages of our simulator we present a thorough analysis of existing network constructor protocols, showcase their behavior and compare their performance to their theoretically computed time complexities as well as to existing protocols.
This is the first experimental study of network constructors and one of the very few in the area of population protocols.

\subsection{Motivation and Contribution}
Our main motivation for this work, was the findings of \cite{MS14,Mi15}, and the performance of the network formation protocols reported  via their theoretical analysis.
Our first goal is a novel simulator capable of processing and executing network constructors protocols with minimal effort from the side of the researcher. 
In particular, we introduce \ACRONYM, a cloud service capable of simulating the performance of network constructors, store the results of the experiments and compare them to other protocols that solve the same problem.
For example, two researchers working on the same problem can upload their work to \ACRONYM and quickly see how it performs in networks of different sizes and under various schedulers, compare their results and understand how they could improve the performance of their work or even easily visualize the network to identify problems with their solution.
This is, from our own experience, very convenient during the development of new protocols.
Results from experiments guide us to further develop our work and improve overall efficiency of the protocols we use. 
More than that, in one specific case our experimental evaluation revealed a certain configuration that seemed highly unlikely in the theoretical analysis that would lead the network to a non terminating condition.
We therefore used simulation traces to adapt the transition map of our protocol and overcome this problematic configuration.

Moreover, population protocols and network constructors are used to model real world systems that are not always expressed by completely random or independent interactions (like the blood flow inside the circulatory system).
As park of our work, we made it possible to extended on different interaction schemes other than the typical random uniform model in order to investigate how existing protocols are affected by time or space locality by selecting the nodes that interact using information about their connectivity, location or interaction history.

In Section~\ref{sec:netcs-model} we provide a more formal definition of network constructors and the main differences from similar models.
Section~\ref{sec:netcs-simulator} offers insights on operation of the \ACRONYM Simulator while Section~\ref{sec:experiments} presents results from the experimental evaluation of various existing protocols.
Finally, in Section~\ref{sec:conclusions} we conclude and give further research directions that are opened by our work.

\section{Network Constructors Model}\label{sec:netcs-model}

The network construction model is strongly inspired by the population protocol model \cite{AADFP06} where nodes have states that change based on the interactions performed.
The mediated population protocol model \cite{MCS11-2} introduced states on the connections between nodes to help solve more complex problems.
The main difference is that in all previous cases the focus was on the computation of functions of some input values and not on network construction.
Another important difference is that network constructors allow the edges to choose between \emph{only two possible states} which is not the case in \cite{MCS11-2}, thus simulating an active or inactive communication link in the final configuration. 
Fields where population protocols and network constructors can be applied include {Algorithmic Self-Assembly}, {Cellular Automata}~\cite{DGRS13}, {Social Networks}, and {Network Formation in Nature}~\cite{Do12}.

A more formal description of network constructors is a set of $n$ processes that are capable of performing local computation (via pairwise interactions) and of forming and deleting connections between them.
In the most general case, a connection between two processes can be in one of a finite number of possible states.
For example, state $0$ could mean that the connection does not exist while state $i \in {1, 2, . . . , k}$, for some finite $k$, that the connection exists and has strength $i$. 
\ACRONYM focuses in the simplest case, called the \texttt{on/off case}, in which, at any time, a connection can either exist or not exist. 
If a connection exists we also say that it is active and if it does not exist we say that it is inactive. 
Initially all connections are inactive and the goal is for the processes, after interacting and activating/deactivating connections for a while, to end up with a desired stable network. 
In the simplest case, the output-network is the one induced by the active connections (\eg a line, a ring, a star or a cycle cover) and it is stable when no connection changes state any more.

The communication model of the system in question is also very minimal, yet highly extensible. 
In particular, processes are inhabitants of an adversarial environment that has total control over the inter-process interactions. 
This environment is modeled by an adversary scheduler that operates in discrete steps selecting in every step a pair of processes that then interact according to the common program. 
This represents very well systems of (not necessarily computational) entities that interact in pairs whenever two of them come sufficiently close to each other. 
When two processes interact, the program takes as input the states of the interacting processes and the state of their connection and outputs a new state for each process and a new state for the connection. 
The only restriction imposed on the scheduler in order to guarantee the constructive power of the model is that it is fair, by which we mean the weak requirement that, at every step, it assigns to every reachable configuration of the system a non-zero probability to occur.
In other words, a fair scheduler cannot forever conceal an always reachable configuration of the system. 

What renders this model interesting is its ability to achieve complex global behavior via a set of notably simple, uniform (\ie with codes that are independent of the size of the system), homogeneous, and cooperative entities.

\section{The Network Constructors Simulator}\label{sec:netcs-simulator}

A lot of work has been devoted in the past years to develop tools that simulate the behavior of computer networks \cite{ns3} and wireless sensor networks \cite{shawn,tossim}. 
Nevertheless, such tools remain mostly domain-specific and extremely hard to modify in order to prove useful in domains that maintain the same principles with computer or sensor networks but introduce their own characteristics.

To the best of our knowledge, most attempts to simulate the behavior of population protocols (the base model) use custom tools developed with minimal re-usability and limited functionality.
Actually, in most papers the analysis of new protocols remains only theoretical and when experimental results are provided little insight is given to the tools used to simulate their behavior. 
As a result, comparing different results is hard or even impossible as there is no common basis for the performance or validity of the experiments.

\ACRONYM is a cloud service designed for simulating network constructors protocols and storing experimental results from the simulations executed.
It is split in three independent components the Experiment Execution Engine, the Experiment Storage System and the Web UI that provides access to the functionality of the two other components.
All three components are implemented in Java and are available on Github\footnote{https://github.com/amaxilat/netcs} under  the  BSD  3-Clause  License.
\ACRONYM is built using well-established cloud technologies like the SpringBoot framework for the cloud services and MongoDB for storing the experimental and user data.

\begin{figure}[h]
\centering
\includegraphics[width=0.9\textwidth]{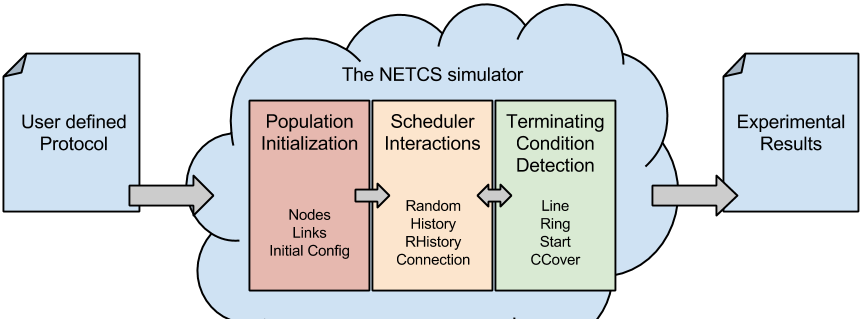}
\caption{The internal architecture of \ACRONYM.}
\label{fig:NETCS:architecure}
\end{figure}

The overall experimenting flow is described in Figure~\ref{fig:NETCS:architecure}.
Users submit their protocols in the service and then they can easily configure a set of experiments by specifying the size of the network, the number of iterations and the schedulers to simulate.
The data provided are forwarded to the Experiment Execution Engine that takes over the executions of the experiments.
While the experiments are being executed, users can view the progress of each experiment and see a visualization of the network at the current state of the execution. 
Also another view is available where various aspects of the experimental results (interactions, success rates, deviation from mean results) are presented graphically.

The execution of the experiments is split in three stages.
The first stage is the initialization of the network.
Nodes are created and the initial configuration of the protocol is applied to the  nodes and links.
The second stage is the execution of multiple interactions between nodes of the network until the expected configuration is reached.
After the network has reached its final state, the third stage of each experiment is about extracting the experimental results concerning information like the number of iterations executed or the states of all nodes in the network.
At this point, the data is forwarded to the Experiment Storage Engine for the data to be saved in the database.

\ACRONYM is capable of simulating network constructors using different schedulers to perform the interactions in each step of the execution.
Thee simplest fair probabilistic scheduler commonly used is the uniform random scheduler that selects in every step independently and uniformly at random a pair of processes to interact from all such pairs.
Adding new schedulers is fairly easy as users simply need to add the code that selects the nodes for each interaction and \ACRONYM will identify it and include it to the system.
As part of this work we present 3 more schedulers apart from the \texttt{Random} uniform.
All of them select the first node (\texttt{A}) of the interaction uniformly and then the second node (\texttt{B}) using a different strategy.
\texttt{History} selects \texttt{B} with a 75\% probability as one of the 50 last nodes \texttt{A} interacted with and as a random node otherwise (25\% probability). 
\texttt{ReverseHistory} uses a (25\%,75\%) probability instead to achive the opposite effect.
Finally \texttt{Connection} selects \texttt{B} from the active links of \texttt{A} with an 80\% probability and uniformly otherwise.
Based on the characteristics of each protocol, different schedulers are expected to affect differently the running times required. 

More detailed information about the internal operations of \ACRONYM is available in the project's wiki on Github.

\section{Experimental Evaluation}\label{sec:experiments}

In the rest of this section we present the experimental evaluation performed on existing network constructors and showcase the results of more than 100000 experiments to validate the operation of \ACRONYM as well as innovative results that extend the  knowledge on the behavior of the protocols themselves.
In some cases, we even present variations that perform better than the existing ones.
Section~\ref{sec:experimentation:conf} presents the evaluation of constructors that generate Lines, Rings, Stars and Cycle Covers and their respective performance for different population sizes and schedulers. Section~\ref{sec:experimentation:size} evaluates on the probabilistic counting problem introduced in \cite{Mi15}.
Our results show how the nodes of a set of distributed processes can estimate the size of the their network using a unique leader or not with an upper bound on execution time.

\subsection{Specific Network Constructions}
\label{sec:experimentation:conf}

For each problem considered in this work we here provide formal definitions, protocols and bounds. 
We also evaluate their performance under 4 different schedulers (\texttt{Random, History, ReverseHistory and Connection}) to investigate if and how much they affect their total execution time. 

\begin{figure}[h!]
\centering
\includegraphics[width=0.8\textwidth]{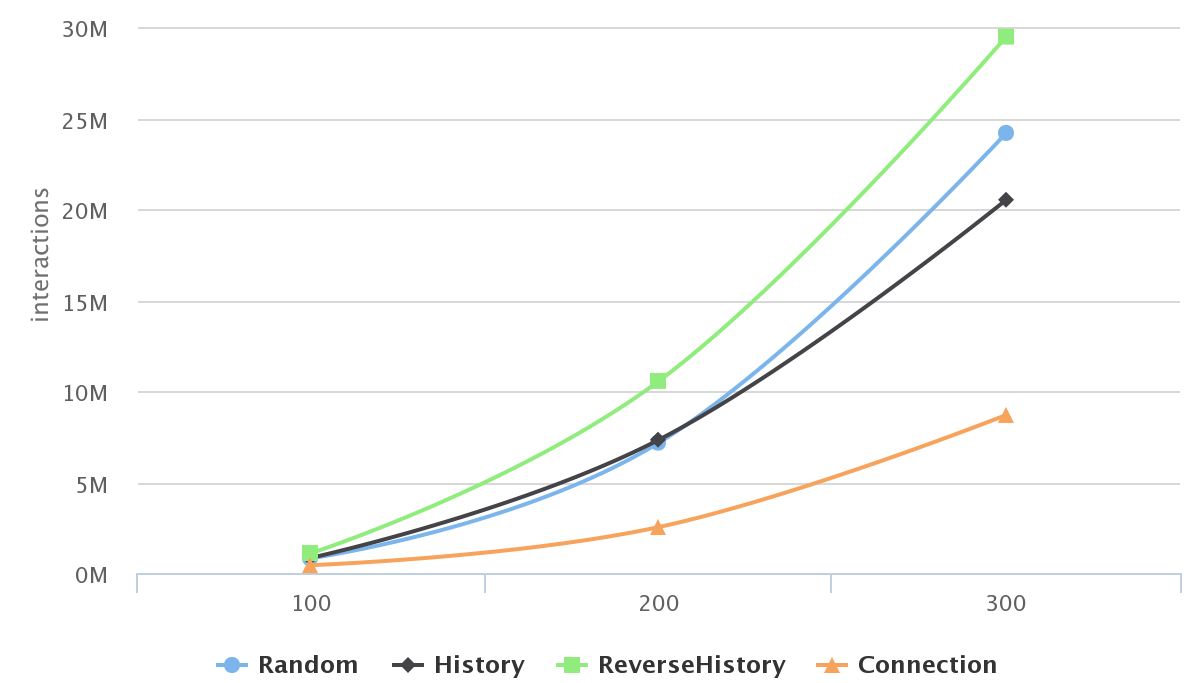}
\includegraphics[width=0.8\textwidth]{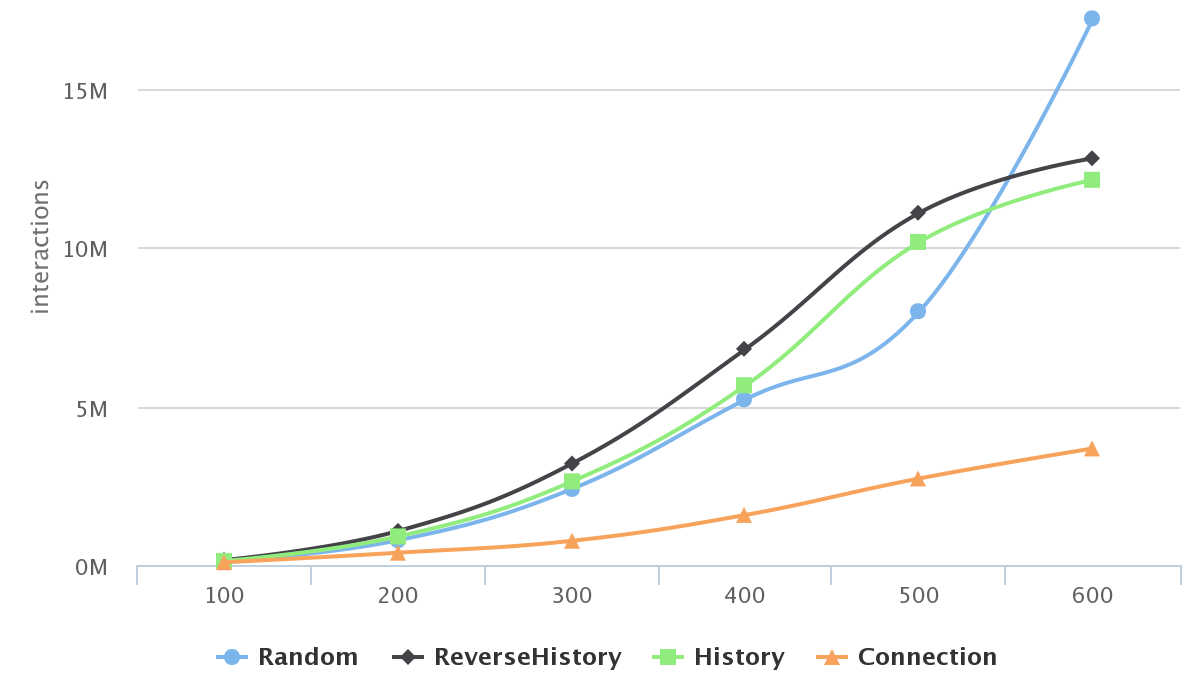}
\caption{The experimental behavior of \textsf{Fast} (up to 300 nodes) and \textsf{Faster} (up to 600 nodes) \gline protocols.}
\label{fig:global-line}
\end{figure}

\subsubsection{\gline} The goal is for the $n$ distributed processes to construct a spanning line, \ie a connected graph in which 2 nodes have degree 1 and $n-2$ nodes have degree 2.
As proven in \cite{MS14} the expected convergence time of any protocol that constructs a spanning line is $\Omega(n^2)$.
We here present the evaluation of two protocols (Fast and Faster GlobalLine).
We begin with a short description of the Protocol~\ref{prot:gline3}:

\floatname{algorithm}{Protocol}
\renewcommand{\algorithmiccomment}[1]{// #1}
\begin{algorithm}[!h]
  \begin{minipage}[b]{0.42\textwidth}
  \caption{\textsf{Fast-\gline}}\label{prot:gline3}
  \begin{algorithmic}
    \medskip
    \State $Q=\{q_0,q_1,q_2,q_2^\prime,l,l^\prime,l^\dprime,f_0,f_1\}$
    \State $\delta$: 
    \begin{align*}
    (q_0,q_0,0)&\ra (q_1,l,1)\\
    (l,q_0,0)&\ra (q_2,l,1)\\
    (l,l,0)&\ra (q_2^\prime,l^\prime,1)\\
    (l^\prime,q_2,1)&\ra (l^\dprime,f_1,0)\\
    (l^\prime,q_1,1)&\ra (l^\dprime,f_0,0)\\
    (l^\dprime,q_2^\prime,1)&\ra (l,q_2,1)\\
    (l,f_0,0)&\ra (q_2,l,1)\\
    (l,f_1,0)&\ra (q_2^\prime,l^\prime,1)
    \phantom{\hspace{10cm}}
    \end{align*}
  \end{algorithmic}
  \end{minipage}
\end{algorithm}

The configuration is always a collection of awake (with an $l$ leader) and sleeping (with an $f_1$ leader) lines and isolated nodes (either awake in $q_0$ or sleeping in $f_0$). 
When two disjoint lines interact, the corresponding leaders play a game in which only one survives. 
The winner grows by one towards the other line and the loser sleeps. 
In particular, when two $l$ leaders interact one of them becomes $l^\prime$ and the other becomes $q_2^\prime$. 
The $l^\prime$ waits to interact with its $q_2$ (or $q_1$) neighbor to convert it to $f_1$ (or $f_0$, resp.) and detach from it, leaving it the endpoint of a sleeping line (or a sleeping isolated node, resp.). 
Then the leader, which is now in $l^{\prime\prime}$, waits to meet again its $q_2^{\prime}$ neighbor to convert it to $q_2$ and update itself to $l$. 
This completes the operation of a line growing one step towards another line and making the other line sleeping. 
A sleeping line cannot increase any more and only loses nodes by lines that are still awake by a similar operation as the one just described. 
A single leader is guaranteed to always win and this occurs quite fast. 
Then the leader makes progress (by one) in most interactions and every such progress is in turn quite fast.
As long as there are at least two awake lines, eventually another line becomes sleeping, so eventually a single line remains awake with all other nodes sleeping (either part of a sleeping line or isolated). 
The protocol ensures that an awake line can always grow towards sleeping nodes (either by stealing them from sleeping lines or by expanding towards isolated nodes), so eventually the unique awake line becomes spanning.

\floatname{algorithm}{Protocol}
\renewcommand{\algorithmiccomment}[1]{// #1}
\begin{algorithm}[!h]
  \begin{minipage}[b]{0.42\textwidth}
  \caption{\textsf{Faster-\gline}}\label{prot:gline4}
  \begin{algorithmic}
    \medskip
    \State $Q=\{q_0,q_1,q_2,q,l,f\}$
    \State $\delta$: 
    \begin{align*}
    (q_0,q_0,0)&\ra (q_1,l,1)\\
    (l,q_0,0)&\ra (q_2,l,1)\\
    (l,q,0)&\ra (q_2,l,1)\\
    (l,l,0)&\ra (l,f,0)\\
    (f,q_2,1)&\ra (q,f,0)\\
    (f,q_1,1)&\ra (q,q,0)
    \phantom{\hspace{10cm}}
    \end{align*}
  \end{algorithmic}
  \end{minipage}
\end{algorithm}

\textsf{FasterGlobalLine} operates similarly but uses some kind of parallel execution:
As in \textsf{FastGlobalLine}, many lines grow in parallel. When the leaders of two lines interact, one of them becomes a follower $f$. The follower starts deactivating its own line, releasing its nodes, while the $l$ that survived does not change its behavior. 
Observe the contrast to the \textsf{FastGlobalLine} protocol: in that protocol sleeping lines could only lose nodes by interacting with awake leaders, while now sleeping lines keep releasing their own nodes to make them available to the awake leaders. 
Eventually, a single $l$ will remain and all other lines will have an $f$. 

As part of this experiment we focus here on proving that the parallel releasing of the nodes of the $f$-lines allows the $l$ leader to be able to rapidly expand towards free nodes.
Also we observe that the description of this protocol is quite simpler than the description of \textsf{FastGlobalLine} yet its performance proves to be a lot better than the first one.
Figure~\ref{fig:global-line} and Table~\ref{tab:global-line} show the performance of the two global line constructors. 
Both, based on the results of \cite{MS14}, have a theoretical time complexity of $O(n^3)$ but as it is clear from our experiments \textsf{FasterGlobalLine} is clearly much faster due to the speedup we described above.

\begin{table}
\caption{Hidden coefficient of the Fast and Faster \gline protocols}
\label{tab:global-line}
\centering
\small
\begin{tabular}{|c|r|r|r|r|r|r|r|r|}
\hline

 &  \multicolumn{4}{c|}{Fast Global Line $O(n^3)$} & \multicolumn{4}{c|}{Faster Global Line $O(n^3)$} \\
   \cline{2-9}
N &Random &History &RHistory &Connection  & Random & RHistory & History & Connection\\
 \hline
100 & 0.82 & 0.84 & 1.10 & 0.43 & 0.12 & 0.16 & 0.13 & 0.09 \\
200 & 0.90 & 0.92 & 1.32 & 0.32 & 0.10 & 0.13 & 0.11 & 0.05 \\
300 & 0.90 & 0.76 & 1.09 & 0.32 & 0.09 & 0.12 & 0.10 & 0.03 \\
400 &      &      &      &      & 0.08 & 0.11 & 0.09 & 0.02 \\
500 &      &      &      &      & 0.06 & 0.09 & 0.08 & 0.02 \\
600 &      &      &      &      & 0.08 & 0.06 & 0.06 & 0.02 \\

\hline
\end{tabular}
\end{table}

\subsubsection{\gstar} 
\label{sec:experimentation:star}
The idea behind the protocol~\cite{MS14} is that nodes may play one of the following two roles during an execution: a \emph{center} (state $c$) or a \emph{peripheral} (state $p$). 
The unique output-stable configuration $C_f$ whose active network is a spanning star, has one center and $n-1$ peripheral nodes, and a $uv$ edge is active iff one of $u,v$ is the center. 
Initially all nodes are centers. When two centers interact one of them remains a center and the other becomes a peripheral. 
No other interactions eliminate a center, which implies that not all centers can be eliminated, and once a center becomes a peripheral it never becomes a center again. 
Due to fairness, eventually all pairs of centers will interact and, as no new centers appear, a single center will prevail. 
Thus from some point on there is a single center and $n-1$ peripheral nodes leading to the construction of a spanning star.

\begin{figure}[h]
\centering
\includegraphics[width=0.8\textwidth]{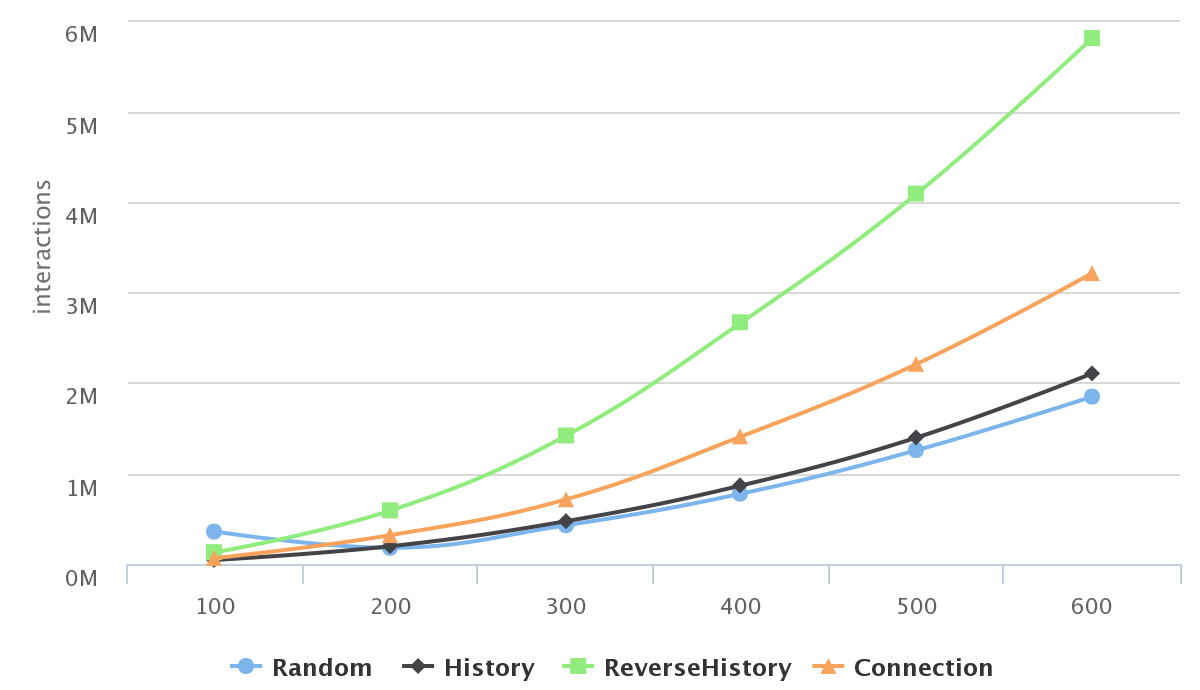}
\caption{The experimental behavior of the \gstar protocol.}
\label{fig:global-star}
\end{figure}
Analysis of the \gstar protocol presented in \cite{MS14} shows that its expected running time is $O(n^2\log{n})$ a bound that is confirmed by our experimental results (over 7000 experiments) presented in Figure~\ref{fig:global-star}.
The coefficients of the big-o notation under all used schedulers are available in Table~\ref{tab:global-star} and remain relatively stable for all of them.
\gstar targets to connect all nodes of the network to the leader node so that they are deactivated by him and the protocols finishes its execution. 
This makes it necessary for the center node to communicate with nodes that are across the whole network and results to a deteriorated performance with the ReverseHistory scheduler while History is slightly worse than Random.
This is also expected as nodes should communicate only a limited number of times before they reach their final state and communicating with already visited nodes does not offer any speedup.
For similar reasons the Connection scheduler has also a small negative impact on the performance of the protocol. 

\begin{table}
\caption{Hidden coefficient of the \ccover and \gstar protocols}
\label{tab:cycle-cover}
\label{tab:global-star}
\centering
\small
\begin{tabular}{|c|r|r|r|r|r|r|r|r|}
\hline
 &  \multicolumn{4}{c|}{Cycle Cover  $O(n^2)$ } & \multicolumn{4}{c|}{Global Star $O(n^2\log{n})$} \\
   \cline{2-9}
N & Random & History & RHistory & Connection &Random &History &RHistory &Connection \\
\hline
100 & 3.93 & 29.56 & 3.26 & 2.72  & 7.77 & 0.98 & 2.83 & 1.43 \\
200 & 0.74 & 6.09 & 2.76 & 2.75   & 0.85 & 0.93 & 2.80 & 1.51 \\
300 & 0.71 & 0.88 & 4.35 & 2.73   & 0.83 & 0.92 & 2.76 & 1.39 \\
400 & 0.70 & 0.91 & 2.98 & 1.81   & 0.81 & 0.91 & 2.78 & 1.47 \\
500 & 0.70 & 0.82 & 2.70 & 3.55   & 0.81 & 0.90 & 2.63 & 1.42 \\
600 & 0.67 & 0.91 & 2.62 & 3.32   & 0.80 & 0.91 & 2.52 & 1.39 \\

\hline
\end{tabular}
\end{table}

    
    
    

\subsubsection{\ccover}
\label{sec:experimentation:ccover} presented in \cite{MS14} operates by preserving the following invariant: 
the degree of a node in state $q_i$, $0\leq i\leq 2$, is $i$. 
All interactions $(q_i,q_j,0)$ with $i,j\in\{0,1\}$ result in $(q_{i+1},q_{j+1},1)$, that is an activation and a corresponding increase in the recorded degrees. 
As a result, as long as there are at least two disconnected nodes with degrees smaller than two, these two nodes can become connected. 
It follows that any component with at least three nodes eventually becomes a cycle and in the final stable configuration there can be at most one component that is not a cycle: either an isolated node, or two nodes connected by an active edge. 
Its expected running time under the uniform random scheduler is $\Theta(n^2)$, and is optimal with regard to time.
Every process in $V_I$ eventually has a degree of 2 and the result is a collection of node-disjoint cycles spanning $V_I$. 

\begin{figure}[h]
\centering
\includegraphics[width=0.8\textwidth]{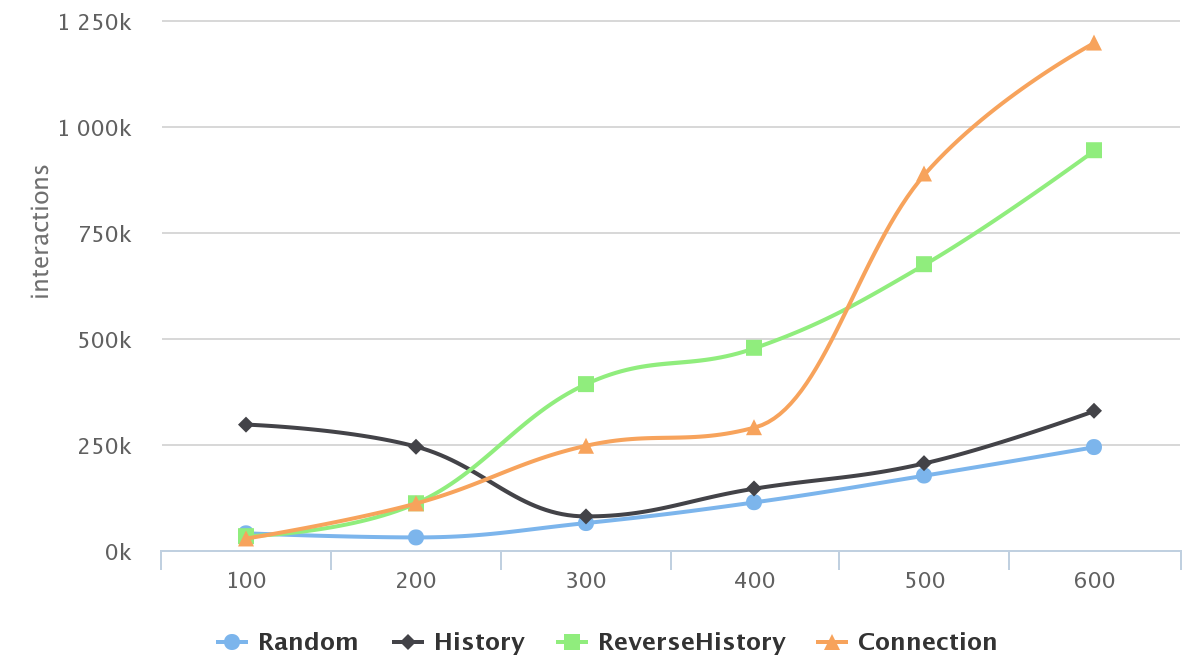}
\caption{The experimental behavior of the \ccover protocol.}
\label{fig:cycle-cover}
\end{figure}

From out experimental evaluation presented in Figure~\ref{fig:cycle-cover} (a total of 5000 simulations) we confirm that the operation of the \ccover protocol is limited by $\Theta(n^2)$ for the \texttt{Random} and \texttt{History} schedulers but deteriorates when we use \texttt{ReverseHistory} or \texttt{Connection}.
This behavior is expected as the protocol does not deactivate any connections of the network during its operation and as a result \texttt{Connection} and \texttt{ReverseHistory} schedulers tend to introduce more interactions that do not create any change to the network.
Table~\ref{tab:cycle-cover} shows the big-o hidden coeficients for the \ccover protocol and confirms the adherence to the $O(n^2)$ complexity under the \texttt{Random} and \texttt{History} schedulers.

\subsubsection{\gring}  was originally presented in \cite{MS14} but from our experimental evaluation we were able to identify some cases where the protocol failed to create an actual ring in the network.
Using the feedback from the experiments we were able to make the required changes and introduce a new state that helps to overcome the problem and generate a global ring in all experiments. 
Furthermore, the current set of experiments act as a first result towards specifying the time complexity of the protocol since there has been no theoretical analysis before.
\gring is a clear example about how schedulers like Connection or History can significantly affect the performance of protocol.


Using the Random scheduler the number of interactions required for the ring to form is extremely high.
Using a scheduler like Connections that favors interactions between nodes that have communicated in the past significantly reduces the number of interactions required and leads to lower execution times.
More information on the execution times required by the \gring protocol are available in Table~\ref{tab:global-ring}.

\begin{table}
\caption{Execution times in millions of interactions for the \gring protocol}
\label{tab:global-ring}
\centering
\small
\begin{tabular}{|c|r|r|r|r|}
\hline

N & Random & Connection & RHistory & History \\
\hline

100 & 4.4  & 0.32 & 7.8 & 4.7 \\
200 & 381  & 1.54 & 20  & 20  \\
300 & 1000 & 5.9  &     &     \\

\hline
\end{tabular}
\end{table}

\subsection{Probabilistic Counting}
\label{sec:experimentation:size}

In this section we focus on the probabilistic counting problem. 
We assume a uniform random scheduler and we want to give protocols that always terminate but still w.h.p. count $n$ correctly (or a satisfactory upper bound on $n$). 
The following is a protocol with a unique leader that solves w.h.p. the counting problem and always terminates \cite{Mi15}.

\noindent\textbf{Counting-Upper-Bound Protocol:} There is initially a unique leader $l$ and all other nodes are in state $q_0$. Assume that $l$ has two $n$-counters in its memory, initially both set to 0. So, the state of $l$ is denoted as $l(r_0,r_1)$, where $r_0$ is the value of the first counter and $r_1$ the value of the second counter, $0\leq r_0,r_1\leq n$. The rules of the protocol are $(l(r_0,r_1),q_0)\ra (l(r_0+1,r_1),q_1)$, $(l(r_0,r_1),q_1)\ra (l(r_0,r_1+1),q_2)$, and $(l(r_0,r_1),\cdot)\ra (halt,\cdot)$ if $r_0=r_1$.

Observe that $r_0$ counts the number of $q_0$s in the population while $r_1$ counts the number of $q_1$s. Initially, there are $n-1$ $q_0$s and no $q_1$s. Whenever $l$ interacts with a $q_0$, $r_0$ increases by 1 and the $q_0$ is converted to $q_1$. Whenever $l$ interacts with a $q_1$, $r_1$ increases by 1 and the $q_1$ is converted to $q_2$. The process terminates when $r_0=r_1$ for the first time. We also give to $r_0$ an initial head start of $b$, where $b$ can be any desired constant. So, initially we have $r_0=b$, $r_1=0$ and $i=\#q_0=n-b-1$, $j=\#q_1=b$ (this can be easily achieved by the protocol). So, we have two competing processes, one counting $q_0$s and the other counting $q_1$s, the first one begins with an initial head start of $b$ and the game ends when the second catches up the first. 
As it was proved in \cite{Mi15}, when this occurs the leader will almost surely have already counted at least half of the nodes.
This is captured in the following theorem:

\begin{theorem}[\cite{Mi15}]\label{the:count-half}
The above protocol halts in every execution. Moreover, if the scheduler is a uniform random one, when this occurs, w.h.p. it holds that $r_0 \geq n/2$.
\end{theorem}

\begin{remark}
For the Counting-Upper-Bound protocol to terminate, it suffices for the leader to meet every other node twice. This takes twice the expected time of a \emph{meet everybody} (cf. \cite{MS14}), thus the expected running time of Counting-Upper-Bound is $O(n^2\log n)$ (interactions).
\end{remark}

We conducted more than 42000 experiments for various sizes ranging from 10 to 1000 nodes and in 99\% of the experiments the nodes terminated after having calculated over 90\% of the total size of the network.
Also the experimentally calculated time to converge is bounded by  $O(n^2\log n)$ with a hidden coefficient of $0.70$ to $0.74$. 

Additional experiments we performed, showed that any given network constructor, where nodes start from a common state, reaches a configuration where every state $q \in Q$ exists in the network with a constant multiplicity of size $\Theta(n)$~(a fact implied by \cite{DotyTCRN14}) and this configuration continuous to exist in the network for $\Theta(n)$ interactions.
In more detail, we observed that the lower the number of states is the longer the configuration described above is detected ($15n$ when $|Q|=4$, $6n$ when $|Q|=5$ and $2n$ when $|Q|=6$).
To validate our result we experimented with over 40 random protocols with $|Q| \in [4,6]$ and network sizes ranging from $100$ to $1000$ nodes.
These results, provide some first experimental evidence that it might be impossible for any protocol to estimate the size of any network without the existence of a unique leader (this is an intriguing theoretical question left open by \cite{Mi15}).

\section{Conclusions and Further Research}
\label{sec:conclusions}
Our experience from implementing and evaluating the performance of \ACRONYM has provided us with a lot of insight on the difficulties of simulating population protocols and variations.
The behavior and requirements of each protocol can be quite different and maintaining a common tool is challenging yet possible from our findings.
Also the usage of such a tool has proven to be extremely useful during all stages of a protocol development.

As part of our future work we would like to focus on the ability to execute simulations in a distributed manner over multiple machines in order to extend the computational platform.
This has proven to be the most difficult part to improve in a single server environment. 
Additionally, we would like to continue extending the basis of the available schedulers and protocols so that we can provide a rich experimental platform for researchers not only in the basis of population protocols and network constructors but in highly dynamic distributed systems and self-assembly systems in general.


\bibliographystyle{splncs03}
\bibliography{main,michail15-arxiv,podc14-full}


\end{document}